\documentclass[twocolumn,superscriptaddress]{revtex4}
\usepackage{graphicx}
\usepackage{epstopdf}
\usepackage{amsmath}
\begin{document}
%\draft
%twocolumn
%[\hsize\textwidth\columnwidth\hsize\csname@twocolumnfalse\endcsname
\draft
\title{Phase diagram of patchy colloids:  towards  empty liquids}
\author{Emanuela Bianchi} 
\affiliation{ {Dipartimento di Fisica and
  INFM-CRS-SMC, Universit\`a di Roma {\em La Sapienza}, Piazzale A. Moro
  2, 00185 Roma, Italy} } 
\author{Julio Largo} \affiliation 
{ {Dipartimento di Fisica,  Universit\`a di Roma {\em La Sapienza}, Piazzale A. Moro
  2, 00185 Roma, Italy} } 
  \author{Piero Tartaglia} 
\affiliation{ {Dipartimento di Fisica and
  INFM-CRS-SMC, Universit\`a di Roma {\em La Sapienza}, Piazzale A. Moro
  2, 00185 Roma, Italy} } 
\author{ Emanuela Zaccarelli}  \affiliation{ {Dipartimento di Fisica and
  INFM-CRS-SOFT, Universit\`a di Roma {\em La Sapienza}, Piazzale A. Moro
  2, 00185 Roma, Italy} } 
\author{  Francesco Sciortino} \affiliation{ {Dipartimento di Fisica and
  INFM-CRS-SOFT, Universit\`a di Roma {\em La Sapienza}, Piazzale A. Moro  2, 00185 Roma, Italy} }

\begin{abstract}
We report theoretical and numerical evaluations of the
phase diagram for patchy colloidal particles of new generation.
We show that the reduction of the number of bonded nearest neighbours 
%has two interesting consequences on the thermodynamics of the system: 
%each of them with potential  applications:
%(i) it 
offers the possibility of generating liquid states (i.e. 
states with temperature $T$ lower than the liquid-gas critical temperature) with a
vanishing occupied packing fraction  ($\phi$), a case which can not be realized with spherically interacting particles. Theoretical results suggest that such reduction is accompanied by 
an increase of the  region of stability of the liquid phase in the ($T$-$\phi$) plane,  possibly favoring the establishment of  homogeneous disordered materials at small $\phi$, i.e. stable equilibrium gels.  

\end{abstract}
%82.70.Dd (colloids) , and 82.70.Gg (gels) 61.20.Ja Computer simulation of liquid structure

\maketitle

The physico-chemical manipulation of colloidal particles
is growing at an incredible pace.  The  large freedom in the control of the inter-particle potential  has made it possible to design  colloidal particles which significantly extend the possibilities offered by atomic systems~\cite{Blaad_03}.   An impressive step further is offered by the newly developed techniques to assemble (and produce with significant yield)  colloidal {\it molecules},   particles decorated on their surface by a predefined number of  attractive sticky spots, i.e. particles with specifically designed shapes and interaction sites~\cite{ 
Manoh_03,
Cho_05,
%Yi_04,Cho_05bis,
Zerro_05,mohovald}. 
These new particles, thanks to the specificity of the built-in interactions, will be able not only to reproduce molecular systems on the  nano and micro scale, but will also show novel  collective behaviors.  
To guide  future applications of patchy colloids, to help designing  
bottom-up strategies in self-assembly~\cite{bottomup,Glotz_04,Glotz_Solomon} and   
to tackle the issue of interplay between dynamic arrest and  crystallisation --- a hot-topic  related for example  to the possibility of nucleating a colloidal diamond crystal structure for photonic applications~\cite{fotonic}
%Klein_05,Lu_,Liddell_03} 
---  it is crucial
to be able to predict  the region in the ($T$ - $\phi$) plane in which clustering, phase separation or even  gelation is expected.

While design and production of patchy colloids is present-day research, unexpectedly theoretical studies of the physical properties of these systems have a longer history, starting in the eighties in the context of the physics of associated liquids~\cite{Kol_87,Nez_89,Nez_90,Sear_96,Monson_98,Ford_04}. These studies, in the attempt to
pin-down the essential features of association, modelled
molecules as  hard-core particles with attractive spots on the
surface,  a realistic description of the recently created   patchy colloidal particles. A thermodynamic perturbation theory  (TPT)
appropriate for  these models was introduced by Wertheim~\cite{Werth1} to describe association under  the hypothesis  that a sticky site on a particle cannot bind simultaneously to two (or more) sites on another particle. Such a condition can be naturally implemented in colloids, due to the relative size of the particle as compared to the range of the sticky  interaction. These old studies
%, despite their limitation due to the limited computational power which did not allow for a carefully checking of the theory in conditions of extended bonding or close to the liquid-gas critical point, 
provide a very valuable starting point for addressing the issue of the phase diagram of this new class of colloids, and in particular of the role of the patches number.

\begin{figure}[t] %  figure placement: here, top, bottom, or page
%   \centering
 \includegraphics[width=8cm, clip=true]{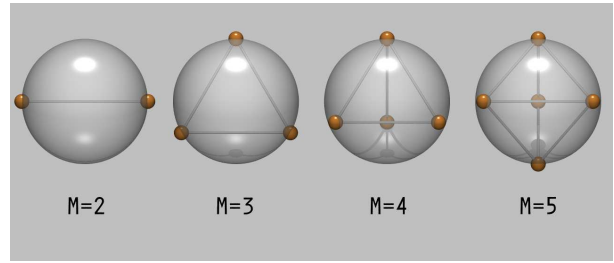}
   \caption{Schematic representation of the location of the square-well interaction sites (centers of the small spheres)  on the surface of the hard-core particle. Sticks between different interaction sites are drawn only to help visualizing the geometry.   
   % The size of the spheres is in scale. When two particles approach each other  in such a way that the gold-spheres partially overlap, an attractive interaction sets-in. 
   }
   \label{fig:345}
\end{figure}

In this Letter we study a system of hard-sphere particles with a small
number $M$ of identical short-ranged square-well attraction sites per
particle (sticky spots), distributed on the surface
with the same geometry as the recently produced patchy colloidal
particles~\cite{Zerro_05}. We identify the number of possible bonds
per particle as the key parameter controlling the location of the
critical point, as opposed to the fraction of surface covered by
attractive patches.
%Such particles are pictorially reproduced in Fig.~\ref{fig:345}.
%, focusing on the effect of  $M$ on the phase diagram 
We present results of extensive numerical simulations of this model
in the grand-canonical ensemble~\cite{frenkelsmith} to evaluate
the location of the critical point of the system in the ($T$ - $\phi$)
plane as a function of $M$. 
We complement the simulation results with 
the evaluation of the region of thermodynamic instability   according to the Wertheim theory~\cite{Werth1,Werth2,Hansennew}.
%,Werth3,Werth4,Werth
Both theory and simulation confirm that,
on decreasing the number of sticky sites, the critical point moves
toward smaller $\phi$ and $T$ values. We note that
while adding to hard-spheres a spherically symmetric attraction creates 
a liquid-gas critical point which shifts toward larger $\phi$ on decreasing
the range of interaction, the opposite trend is presented here when  
 the number of interacting sites  is decreased.   Simulation and theory also provide evidence that for binary mixtures of particles with two and three sticky spots (where
$\langle M \rangle $, the average $M$ per particles, can be varied
continuously down to two by changing the relative concentration of the two species) the critical point shifts continuously
toward vanishing $\phi$. This makes it possible to realize equilibrium
liquid states with arbitrary small $\phi$ ({\it empty liquids}), a
case which can not be realized via spherical potentials.

We focus on a system of  hard-sphere particles (of diameter $\sigma$, the unit of length) whose surface is decorated by $M$  sites (see Fig.~\ref{fig:345}), which we  collectively label $\Gamma$. The  interaction $V({\bf 1,2})$  between particles {\bf 1} and  {\bf 2} is 
\begin{equation}
\label{eqn:xxx}
V({\bf 1,2})=V_{HS}({\bf r_{12}})+\sum_{A \epsilon \Gamma}\sum_{B \epsilon \Gamma} V^{AB}_{W}({\mathbf r}_{_{AB}})
\end{equation}
where  the individual sites are denoted by capital letters, $V_{HS}$ is
the hard-sphere potential, $V^{AB}_{W}(x)$ is a well interaction (of depth $-u_0$ for
$x \leq  \delta$, 0 otherwise) and  ${\bf r_{12}}$ and ${\mathbf r}_{_{AB}}$ are respectively the vectors joining the particle-particle and the site-site
centers\cite{squarewell}. Geometric considerations for a  three touching spheres configuration show that the choice $\delta=0.5(\sqrt{5-2\sqrt{3}}-1)  \approx 0.119$ guarantees that each site is engaged at most in one bond.  With this choice of $\delta$, $M$ is also the maximum number of bonds per particle. Temperature is measured in units of $u_0$ (i.e. Boltzmann constant $k_B=1$).

To locate the critical point, we perform grand-canonical Monte Carlo (GCMC)  simulations
and histogram re-weighting~\cite{Wilding_96} for $M=5$, $4$ and $3$ and for binary mixtures of particles with $M=3$ (fraction $\alpha$) and $M=2$  (fraction $1-\alpha$) at five different compositions, down to $\langle M \rangle\equiv 3\alpha+2(1-\alpha) =2.43$.
We implement MC steps composed each by  500 random attempts to rotate and translate a random particle and  one attempt to insert or delete a particle.   On decreasing 
 $\langle M \rangle$,  numerical simulations become particularly time-consuming, since the probability of breaking  a bond $\sim e^{1/T}$ becomes progressively small.  To improve statistics, we average over 15-20 independent MC realizations. 
Each of the  simulations lasts more than $10^{6}$ MC steps. After choosing 
the box size, the $T$ and the chemical potential $\mu$ of the particle(s), the GCMC simulation evolves the system toward the corresponding equilibrium density.  If  
$T$ and $\mu$ correspond to the critical point values, the number of particles $N$   and the potential energy $E$ of the simulated system  show ample fluctuations between  two different values.    The linear combination  $x \sim  N+s E$  (where $s$ is named field mixing parameter) 
plays the role of order parameter of the transition. At the critical point, its fluctuations are found to follow a known universal distribution, i.e. (apart from a scaling factor) the same that characterises the fluctuation of the magnetisation in the Ising model~\cite{Wilding_96}.   Recent applications of this
method to soft matter can be found in Ref.~\cite{Miller_03,puertas,horbach}.

%In addition, we have studied boxes of different size to minimize finite size effects, which are progressively important  on decreasing $M$ due to the small $\phi$ of the gas-like fluctuations.  
%Larger simulations requires significant computational resources, due to the size dependence of the free-energy barrier separating the gas and liquid phases.  

\begin{figure}[t] %  figure placement: here, top, bottom, or page
%   \centering
 \includegraphics[width=8.0cm, clip=true]{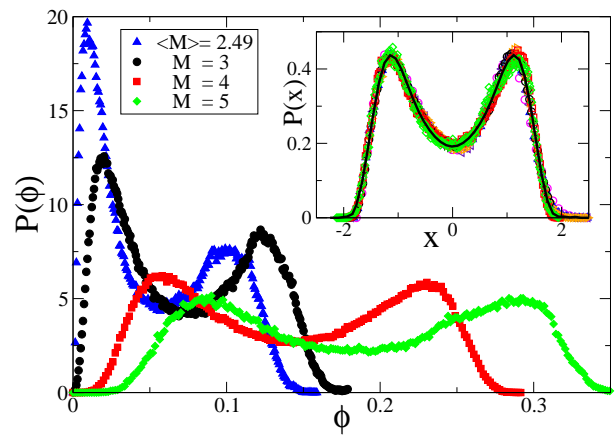}
   \caption{Density fluctuations distribution $P(\phi)$  in the GCMC at
   the critical point for four of the studied $M$ values. The inset shows $P(x)$
   for all studied cases, compared with the expected distribution (full line) for system
   at the critical point  of the Ising universality 
   class~\protect\cite{Wilding_96}.  
   }
\label{fig:onecomponent}
\end{figure}

Fig.~\ref{fig:onecomponent} shows the resulting density fluctuations 
distribution $P(\phi)$ at the estimated critical temperature $T_c$ and critical  chemical potential(s) $\mu_c$
for several $M$ values\cite{bimodal}. The distributions, whose average is the critical packing fraction $\phi_c$, shift to the left on
decreasing $M$ and become more and more asymmetric,
signalling the progressive increasing role of the mixing field.  In the inset, 
 the calculated fluctuations of $x$, $P(x)$, are compared with the expected fluctuations for systems in the Ising universality class~\cite{Wilding_96}  to provide evidence that (i)
 the critical point has been properly located; (ii) the transition belongs to the Ising universality class in all studied cases.
  The resulting critical parameters  are reported in Table~I. Data show a clear monotonic trend toward
 decreasing $T_c$ and  $\phi_c$ on decreasing $M$. 

Differently from the $\phi_c$-scale, which is essentially controlled by $M$, $T_c$ depends 
on the attractive well width. Experimentally,
 values of $u_0/k_BT$ comparable to the ones reported in Table~I can be realized by modifying the physical properties (size, polarizability, charge, hydrophobicity)  of the patches~\cite{Cho_05,Zerro_05,mohovald}
or by functionalizing  the surface of the particle with specific molecules~\cite{dna,weitz}.

 One interesting observation steaming from these results is that reduction of $M$ makes it possible to shift $\phi_c$ to values smaller that $\phi=0.13$, which is the lowest $\phi_c$ possible for attractive spherical potentials.  Indeed for spherical square well potentials   
$0.13<\phi_c<0.27 $, the two limits being provided by the Van der Waals 
(in which repulsion is modelled by the  Carnahan-Starling expression) and 
by the Baxter~\cite{Miller_03} models, respectively with infinite and 
infinitesimal interaction range.   We also note that results  are consistent with those based on a toy model where an ad-hoc constraint was added to limit valency~\cite{Zacca1} and also with previous studies of particles interacting with non-spherical potentials~\cite{Sear_99,Kern_03}.

%awk -F "&" '{print $3,"&",$1,"&",$2,"&",$4,"&",$5,"&",$6}' junk     
\begin{table}[htdp]
\begin{center}
\begin{tabular}{|c|c|c|c|c|c|c|}
 \hline 
   $\langle M \rangle$ &   $T_c$  &  $\phi_c$  &  $\mu_c^1$  &  $\mu_c^2$  &  $s$  & $L$\\
    \hline 
2.43 &     0.076  &  0.036   &  -0.682 &  -0.492  &  0.70 & 9\\
 2.49  &       0.079  &  0.045  & -0.646 &  -0.483   & 0.64 & 9\\
 2.56 &         0.082  &  0.052 &  -0.611 &  -0.478   & 0.57& 9\\
2.64 &    0.084  & 0.055   &  -0.583  & -0.482  &  0.57& 9\\
2.72 &      0.087  & 0.059   &  -0.552  &  -0.493  &  0.52& 9\\
 3 &          0.094  & 0.070  &  -0.471   & - &  0.46   & 9 \\
4  &            0.118  & 0.140  &  -0.418 &  - &   0.08   & 7  \\
 5  &             0.132  &  0.185  &  -0.410  &  -  &  0      & 7    \\    
  \hline 
\end{tabular}
\end{center}
\label{table}
\caption{Values of the relevant parameters at the critical point. In the one-component case ($M=3,4,5$), $\mu_c^1$ is the critical chemical potential (in units of $u_0$). In the case of the mixture,
$\mu_c^1$ ($\mu_c^2$) is the critical chemical potential of $M=3$ ($M=2$) particles. $L$ indicates the largest box size studied.}
\end{table}%

 Visual inspection of the configurations for small $\langle M \rangle$ shows that the system
is composed by  chains of two-coordinated particles providing a link between the
three-coordinated particles, effectively re-normalizing the bonding
distance between the $M=3$ particles. On adding more $M=2$ particles, the bonding distance between $M=3$ particles increases, generating smaller and smaller $\phi_c$.

\begin{figure}[t] %  figure placement: here, top, bottom, or page
%   \centering
 \includegraphics[width=8.5cm, clip=true]{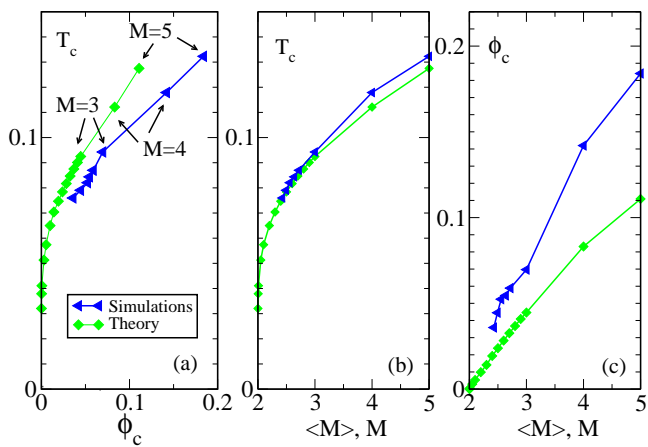}
   \caption{Comparison  between theoretical and numerical results for
   patchy particles with different number of sticky spots. Panel (a) shows the location of the points in the ($T$ - $\phi$) plane.  Panels (b) and (c) compare respectively the  $M$ dependence for $T_c$ and $\phi_c$. 
   }
   \label{fig:tcphic}
\end{figure}

To extend the numerical results beyond the point where it is currently possible
to properly perform GCMC  (at the lowest $\langle M \rangle$  each calculation   of the 20 studied samples requires about 1 month of CPU time on a 3.1 GHz processor) and to complement the numerical
results, we solve the first-order Wertheim TPT~\cite{Werth1,Werth2,Hansennew} for the same model (Eq.~\ref{eqn:xxx}). The theory can be applied both to one component systems ($M=3,4,5$) and  to binary mixtures ( $ \langle M \rangle$ spans continuously the region from $M=2$ ---where no critical point is present--- to $M=3$)~\cite{Chap2}.

In TPT, the free energy of the system is written as the  Helmholtz HS reference free energy $A_{HS}$ plus a  bond contribution $A_{bond}$, which derives by a summation over certain classes of relevant graphs in the Mayer expansion~\cite{Hansennew}.   The fundamental assumption  is that the conditions of steric incompatibilities are satisfied:  (i) no sites can be engaged in more than one bond; (ii) no pair of molecules can be double bonded. The chosen $\delta$ guarantees that the steric incompatibilities are satisfied in the present model.  In the more transparent (but equivalent) formulation of Ref.~\cite{Chap1} $A_{bond}$ is written as
\begin{equation}
\label{eqn:FreeEnergyBond}
\frac{\beta A_{bond}}{N}=\sum_{A \epsilon \Gamma}\left( \ln{X_{A}} - \frac{X_{A}}{2} \right) + \frac{1}{2}M.
\end{equation}
Here $X_{A}$ is the fraction  
of sites A that are not bonded. 
%Here $X_{A}$ is the fraction of molecules $not$ bonded to site $A$.
%The sum in (\ref{eqn:FreeEnergyBond}) runs over all  $M$ sites $A$ in the set $\Gamma$.  
The $X_{A}$s are obtained from the mass-action equation
\begin{equation}\label{eqn:Xa}
X_{A}=\frac{1}{1 + \displaystyle \sum_{B \epsilon \Gamma} \rho X_{B} \Delta_{AB}}
\end{equation}
where $\rho=N/V$ is the total number density and $\Delta_{AB}$ is defined by
\begin{equation}\label{eqn:Deltabis}
\Delta_{AB}= 4 \pi \displaystyle \int{g_{HS}(r_{12})\langle f_{AB}(12)\rangle _{\omega_{1},\omega_{2}} r_{12}^{2} d r_{12}}.
\end{equation}
Here $g_{HS}(12)$ is the reference HS fluid pair correlation function, the Mayer $f$-function is $f_{AB}(12)=exp(-V_{W}^{AB}({\mathbf r}_{AB})/k_{B}T)-1$, and   $\langle f_{AB}(r_{12})\rangle _{\omega_{1},\omega_{2}}$\cite{Werth5}  represents an angular  average over all orientations of molecules 1 and 2 at fixed relative distance $r_{12}$.

The evaluation of $\Delta_{AB}$  requires an expression for
$g_{HS}(r_{12})$  in the range where bonding occurs.  
We have used the linear approximation~\cite{Nez_90} 
\begin{equation}\label{eqn:ghsr}
g_{HS}(r)= \frac{1-0.5 \phi}{(1- \phi)^3}-\frac{9}{2}\frac{\phi (1+\phi)}{(1- \phi)^3} (r-1)\end{equation}
which provides the correct Carnahan-Starling~\cite{CS_69} value at contact.

To locate the critical point, we calculate the equation of state $P(V,T) \equiv 
-{\partial (A_{HS}+A_{bond})}/{\partial V}_T$ 
and search for the
$T$ and $\phi$ value at which both the first and the second volume  ($V$) derivative of the
pressure ($P$) along isotherms vanish.  
Fig.~\ref{fig:tcphic} shows a quantitative comparison of the numerical and theoretical estimates for the critical parameters $T_c$ and $\phi_c$.
Theory predicts quite accurately $T_c$ but slightly underestimates
$\phi_c$, nevertheless clearly confirms the $M$ dependence of the two
quantities.   The overall agreement between Wertheim
theory and simulations reinforces our confidence in the theoretical
predictions and  supports the possibility that on further decreasing $\langle M \rangle$,
a critical point at vanishing $\phi$ can be generated.

\begin{figure}[t] %  figure placement: here, top, bottom, or page
%   \centering
 \includegraphics[width=7.2cm, clip=true]{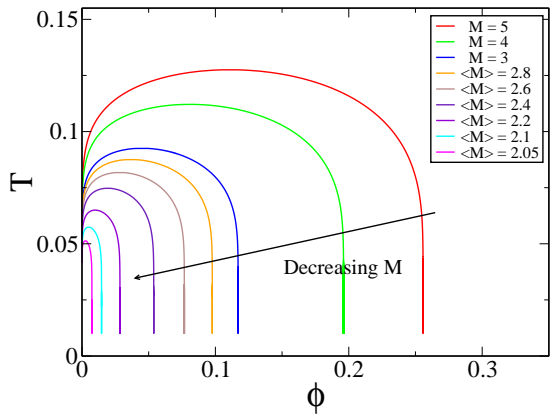}
   \caption{Spinodal curves calculated according to TPT for the studied patchy particles for several $M$ and $\langle M \rangle$ values.  
  }
   \label{fig:pdW}
\end{figure}

TPT allows us also to evaluate the locus of points where $\partial P/\partial V_{T}=0$, which provide (at mean field level) the spinodal locus.   The predicted spinodal lines in the $(T$ - $\phi)$ plane  for several $M$ values  are shown in Fig.~\ref{fig:pdW}.  On decreasing $M$  also the liquid spinodal boundary   moves to lower $\phi$ values, suggesting that the region of stability of the liquid phase is progressively enhanced. It will be desirable to investigate the structural and dynamical properties of such empty liquids by experimental and numerical work on patchy colloidal  particles. 

We note that our predictions are relevant to a larger class of functionalized particles, when particle-particle interaction is selective and limited in number.  Very new materials belonging to this class are the recently synthesised  DNA-coated particles~\cite{dna}. In this case, $M$ can be varied by controlling the number of strands and the attractive strength can be reversibly tuned by varying the length of the strands. Ratios of $u_0/k_BT$ comparable to the ones discussed here can be realized. Again, the phase diagram of these new materials has not been experimentally  measured yet and we hope our work will provide a guideline.  
 
For  particles interacting with attractive spherical potentials, phase separation always destabilises the formation of a homogeneous arrested system at low $T$.  Instead,
it is foreseeable that, with small $\langle M \rangle$ patchy particles,  disordered  states
 in which particles are interconnected in a persistent gel network can be reached at low $T$ without encountering phase separation. Indeed at such low $T$,  the bond-lifetime will  become comparable to the experimental observation time. Under these conditions,  a dynamic arrest phenomenon at small $\phi$  could take place. It would be possible to approach dynamic arrest continuously from equilibrium and to generate a state of matter  as close as possible to an ideal gel~\cite{Sciort_04}.

%\begin{equation}\label{eqn:Mayer} \langle f_{AB}(12)\rangle _{\omega_{1},\omega_{2}}^{M=1} = [(\varepsilon_{AB}/k_{B}T)-1]\frac{(2r_{c} - 2d_{site} + r_{12})(r_{c} + 2d_{site} - r_{12})^2}{24 d_{site}^2 r_{12}}.\end{equation}
 %In agreement  with Coniglio-Klein~\cite{coniglio} description of critical phenomena in term of percolative concepts, the liquid phase is characterized by a percolating transient network of connected particles, a feature which we confirm by analyzing the connectivity of the simulated configurations. Indeed, by analyzing the numerical configurations, we find  that close to the critical point, the average number of bonds per particle is   $\approx 2.3-2.5$ a number which just allows for the establishment of a full percolating network. It is foreseeable that  due to the progressive decrease of $T_c$, the lifetime of these bonds will become longer and longer, giving rise to low $\phi$  percolating long-living structures, with a progressive shift of the viscoelastic cross-over, a feature observed in all liquids, to lower and lower frequency.  

%Zacca1, Zacca2, Moreno_05, Attr_Rep}.   
The study of the structural and dynamic properties of these  low $\langle M \rangle$  equilibrium systems will hopefully help developing a unified picture of other interesting network formation phenomena taking place at low $\phi$~\cite{2000JPCM12R411T,safrannew,delgado,advances}.

We acknowledge support from MIUR-Firb, MIUR-Prin and
MCRTN-CT-2003-504712.  We thank  K. Binder, D. Frenkel and J. Horbach for helpful discussions.

%\cite{Wilding_96, Kern_03, Sear_99, DeMichele_05, Glotz_04, Zhang_04, Zhang_03, Zhang_05, Glotz_Solomon}

\bibliographystyle{apsrev}
\bibliography{./biblio_patchy.bib}

\end{document}